\documentclass[%
 reprint,
superscriptaddress,
 amsmath,amssymb,
 prl,
]{revtex4-2}

\usepackage{ulem}
\usepackage{graphicx}
\usepackage{color} 
\usepackage{dcolumn}
\usepackage{bm}
\usepackage{lipsum}  
\usepackage{amsmath}
\usepackage{dashrule}

\begin{document}

\title{Condensation and activator/repressor control of a transcription-regulated biomolecular liquid}%


\author{Sam Wilken}
\affiliation{%
Physics Department, University of California Santa Barbara, Santa Barbara, CA 93106  USA
}%
\affiliation{%
Materials Department, University of California Santa Barbara, Santa Barbara, CA 93106  USA
}%
\author{Gabrielle R. Abraham}
\affiliation{%
Physics Department, University of California Santa Barbara, Santa Barbara, CA 93106  USA
}%
\author{Omar A. Saleh}
\affiliation{%
Physics Department, University of California Santa Barbara, Santa Barbara, CA 93106  USA
}%
\affiliation{%
Materials Department, University of California Santa Barbara, Santa Barbara, CA 93106  USA
}%

\date{\today}

\begin{abstract}
Cells operate in part by compartmentalizing chemical reactions. 
For example, recent work has shown that chromatin, the material that contains the cell's genome, can auto-regulate its structure by utilizing reaction products (proteins, RNA) to compartmentalize biomolecules via liquid-liquid phase separation (LLPS).
Here, we develop a model biomolecular system that permits quantitative investigation of such dynamics, particularly by coupling a phase-separating system of DNA nanostars to an {\it in vitro} transcription reaction. 
The DNA nanostars' sequence is designed such that they self-assemble into liquid droplets only in the presence of a transcribed single-stranded RNA linker. 
We find that nanostar droplets form with a substantial delay and non-linear response to the kinetics of RNA synthesis.
In addition, we utilize the compartments generated by the phase-separation process to engineer an activator/repressor network, where the formation of droplets is activated by the transcription reaction, and then droplets suppress the transcription reaction by segregating transcription components inside droplets.  
Our work on transcription-driven liquid-liquid phase separation constitutes a robust and programmable platform to explore non-equilibrium reaction-phase transition dynamics and could also provide a foundation to understand the dynamics of transcriptional condensate assembly in cells.
\end{abstract}

\maketitle

\section{Introduction}

Growing experimental evidence~\cite{serizay2018genome,dekker20163d} suggests that eukaryotic genetic material can be organized spatially into membraneless liquid compartments. 
The existence, structure, and properties of these compartments are related both to the level of transcriptional activity~\cite{rada2018forces,cook2018transcription,ulianov2016active} and to phase separation processes in which a liquid mixture of proteins and RNA bathe the genomic DNA~\cite{brangwynne2011active,erdel2018formation,rowley2018organizational}.
Phase transitions differ strongly from classic biomolecular interactions: they exhibit discontinuous responses to changes in the solution conditions and display nonlinear dynamics (e.g. nucleation) due to the highly cooperative nature of the phenomenon~\cite{abraham1974homogeneous,shimobayashi2021nucleation,wilken2024nucleation}.
There has been much recent speculation that chromatin, the material that contains the cell's genome, auto-regulates its structure by utilizing reaction products (proteins, RNA), at least in part, to compartmentalize biomolecules via a liquid-liquid phase transition~\cite{gibson2019organization,wei2020nucleated,strom2024interplay,stortz2024transcriptional}.
Thus, a fundamentally new perspective on cellular function is emerging, in which biomolecular components can be organized in ways that modify the reaction kinetics by which they are produced.

\begin{figure}
\includegraphics[width=8.5cm]{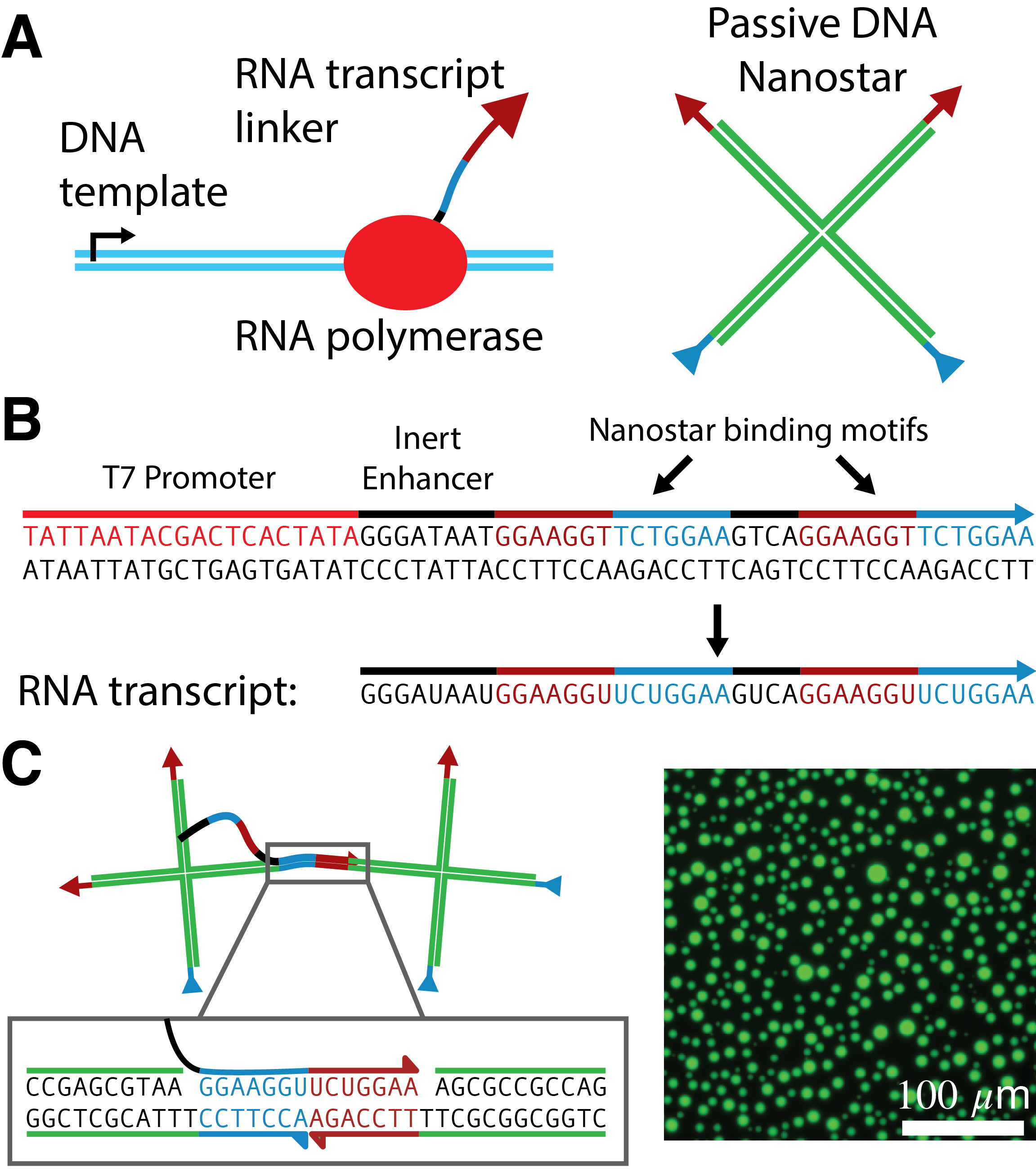}
\caption{\label{fig:nsdesign}
Design scheme of transcription-coupled nanostar phase separation.
{\bf A} RNA polymerase (red) binds to a double-stranded DNA template (cyan) and synthesizes a single-stranded nanostar-binding RNA.
Passive DNA nanostars contain two orthogonal sticky-end binding sequences, allowing for RNA-based linking of two nanostars.
{\bf B} The double-stranded transcription template contains a T7 promoter sequence, an inert enhancer that limits abortive transcripts~\cite{conrad2020maximizing}, and two nanostar-binding motifs to increase RNA synthesis efficiency.
{\bf C} Two sticky ends (red and blue) bind specifically to the transcribed RNA linker, joining nanostars.
Right: fluorescent microscopy image of droplets that result from the mixture of nanostars ($c_{NS} = 28~\mu$M) and purified RNA linkers ($c_{RNA} = 56~\mu$M), incubated at $T=37^\circ$C for 24 hours, in {\it in vitro} transcription buffer conditions excluding RNA polymerase. 
}
\end{figure}

The complexity of living systems suggests that the investigation of such dynamics would be assisted by the development of controlled model systems.
To investigate the interplay between biomolecular reaction kinetics and liquid-liquid phase separation (LLPS), we perform experiments investigating the coupling of transcription reactions to a model phase-separating system of DNA nanoparticles (called `nanostars').
DNA nanostars are roughly 10 nm, finite-valence, self-assembled DNA particles consisting of four double-stranded ``arms", each decorated with a single-stranded sticky end~\cite{biffiPhaseBehaviorCritical2013,abraham2024nucleic,starr2006model,jeon2018salt,conrad2019increasing,sato2020sequence,agarwal2022growth}.
DNA's compatibility with RNA suggests that the nanostar phase transition can be controlled by transcription.



The material properties of the nanostar liquid resemble the chromatin environment.
Nanostar liquids display remarkably low surface tension ($10^{-7} - 10^{-5} N/m$)~\cite{jeon2018salt,sato2023sequence} relative to molecular liquids, but their surface tension values are comparable to estimates of biological condensates ($10^{-7} - 10^{-3} N/m$~\cite{wang2021surface}).
Additionally, the nanostar liquid is a biomolecular example of an `empty liquid'~\cite{bianchi2006phase} due to its open, finite-valence structure ($\approx1\%$ DNA by volume)~\cite{biffiPhaseBehaviorCritical2013,jeon2018salt,conrad2019increasing} which is comparable to the density of DNA in the nucleus ($1-2\%$~\cite{halverson2014melt}). 
Further, the nanostar liquid mesh size, i.e. the characteristic scale of internal voids, has been estimated to be $\approx 8$ nm, as measured by neutral polymer partitioning~\cite{nguyen2019length}, which is similar to estimates of interphase chromatin mesh size ($\approx$10~nm~\cite{wachsmuth2008genome}). 
The physical similarities between nanostar liquids and the {\it in vivo} chromatin environment suggest the potential of DNA nanostars to be utilized as an {\it in vitro} chromatin mimic.

Here, we couple DNA nanostar condensation to a model genetic reaction by designing nanostars that can only be cross-linked by a transcribed RNA molecule. 
We find transcription-driven DNA nanostar condensation dynamics are characteristically non-equilibrium: the condensed volume displays a significant delay and strong non-linear response to synthesized RNA.
We then show that biomolecular partitioning into the droplets can be used to create an activator/repressor network, where {\it in vitro} transcription can control, and be controlled by, the nanostar LLPS process.
This is achieved by engineering a biomolecular spatial segregation scheme in which the droplets contain the genetic template but exclude the RNA polymerase, thus suppressing transcription.

The ability to drive biomolecular condensation with transcription reactions and design auto-regulating transcription-LLPS loops provides a clear method to control reaction pathways in synthetic cells. 
This approach also constitutes a well-defined model system to explore new non-equilibrium dynamical behaviors that depend on the spatial organization of biomolecules, such as cycles of droplet growth and division, boundary effects, and size regulation.

\section{RNA-linked nanostar design}

\begin{figure}
\includegraphics[width=8cm]{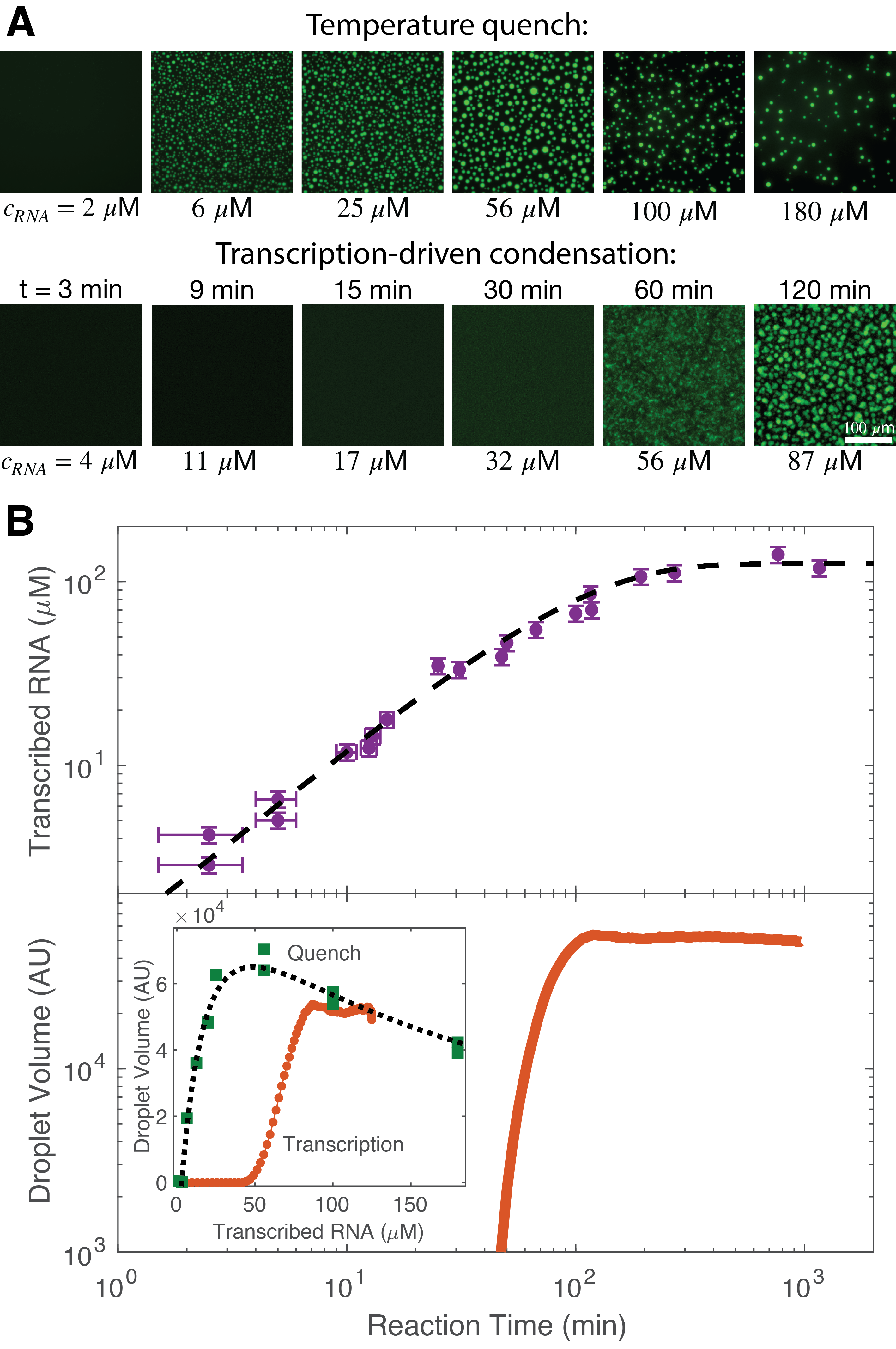}
\caption{\label{fig:nsdynamics} 
Reaction-condensation dynamics of RNA-linked nanostars. 
({\bf A}) For temperature quenches, purified RNA linker/nanostar samples are prepared at different RNA concentrations $c_{RNA}$ and constant nanostar concentration $c_{NS} = 28~\mu$M, and allowed to incubate for 24 hours at $T = 37^\circ$C.
The total droplet volume increases proportionally to RNA linker concentration for $c_{RNA} \lesssim 2c_{NS}$, but then decreases for large $c_{RNA} \gtrsim 2c_{NS}$ (see Inset of {\bf B}). Quench droplet volume is fit well to the functional form $V(c_{RNA}) = V_{max}c_{RNA}/(1+c_{RNA}/2c_{NS})^2$ (dotted line). 
In contrast, the {\it in situ} reaction only produces nanostar droplets after a significant delay ($\approx 45$~min). 
({\bf B}) RNA linkers (purple) are produced at a constant rate before ribonucleotides are depleted and $c_{RNA}$ reaches a plateau (dashed line is fit to $c_{RNA}(t) = c_\infty (1 - e^{-t/\tau_{RNA}})$ with fit parameters $c_{\infty} = 125 \pm 5~\mu$M and $\tau_{RNA} = 120 \pm 10~$min). 
At late stages, the droplet volume (orange) approaches that of the temperature quenches (Inset).
Reactions conducted with $c_{NS} = 28~\mu$M and RNAP concentration $c_{RNAP} = 80~$nM.
RNA quantification includes data from 5 distinct reactions, and the droplet volume trace is a representative curve from one reaction; replicates showed near-identical behavior.
}
\end{figure}

In order to trigger nanostar phase separation with transcribed single-stranded RNA, we use a composite nanostar design that exploits the polarity of nucleic acid hybridization.
Each nanostar contains two different arms with distinct binding sequences (5'-ACCTTCC-Arm) and (Arm-TTCCAGA-3'), permitting single-stranded RNA to link two different nanostars, as shown in Fig.~\ref{fig:nsdesign}. 
Nanostar overhanging sequences are designed to be orthogonal to prevent self-binding in the absence of RNA, which we confirm in Fig.~\ref{fig:nsdynamics}A.

RNA linkers are produced via run-off synthesis of {\it in vitro} transcription reactions performed with T7 RNA polymerase (NEB)~\cite{beckert2011synthesis}.
The polymerase binds to a double-stranded DNA template containing the 17-base T7 promoter sequence and synthesizes a single-stranded RNA copy of the template sequence after the promoter.
We design the template sequence to optimize transcriptional efficiency and DNA nanostar compatibility.
The first 8 bases after the promoter sequence strongly influence transcription efficiency, so we chose a post-promoter `enhancer' sequence, GGGATAAT, previously found to be highly efficient~\cite{conrad2020maximizing} to maximize the transcription rate and minimize incomplete RNA synthesis; however, some incomplete transcription (in the range of 2-8 bases) still occurs~\cite{hsu2009monitoring}.
Additionally, we designed the template to synthesize two binding motifs for each RNA linker strand, separated by an inert 4-base (GTCA) spacer for increased transcription speed and efficiency (see Fig.~\ref{fig:nsdesign}). 
We conduct transcription reactions with a ribonucleotide stoichiometry that matches the transcribed RNA sequence (40\% rGTP, 30\% rATP, 22\% rUTP, 8\% rCTP)~\cite{beckert2011synthesis}.

\begin{figure*}
\includegraphics[width=17cm]{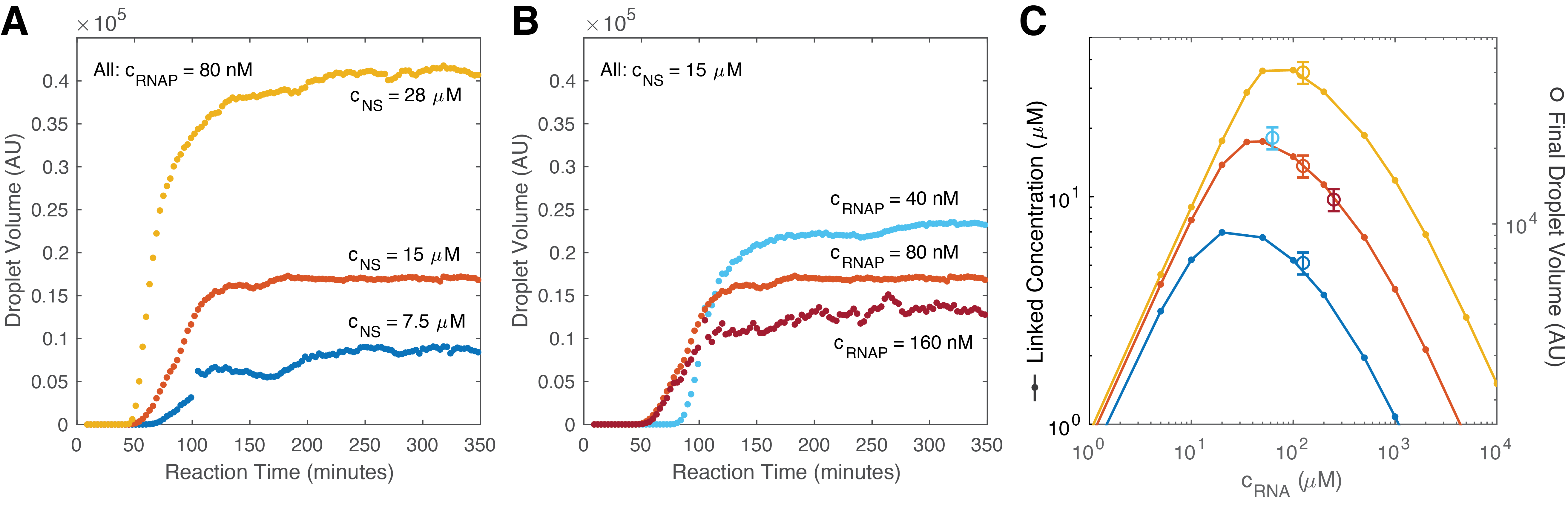}
\caption{\label{fig:ns_rnap_var} 
Nanostar droplet formation dynamics in varied nanostar $c_{NS}$ and RNA polymerase (RNAP) $c_{RNAP}$ concentrations.
{\bf A} For increasing $c_{NS}$ (and constant $c_{RNAP} = 80~$nM), droplets form earlier for larger nanostar concentration, a behavior consistent with the diffusion-coalescence mechanism that governs near-equilibrium droplet growth. 
The final condensed droplet volume is proportional to $c_{NS}$, as expected from the equilibrium lever rule. 
{\bf B} Condensation reactions at varied $c_{RNAP}$ (and constant $c_{NS} = 15~\mu$M) show that the onset of droplet formation does not depend strongly on $c_{RNAP}$ (and therefore RNA-linker concentration $c_{RNA}$, $c_{RNA} \propto c_{RNAP}$). However, the final droplet volume decreases with increasing $c_{RNA}$. 
{\bf C} NUPACK calculated RNA-nanostar construct concentration as a function of $c_{RNA}$ shows non-monotonic behavior consistent with temperature quenches in Fig.~\ref{fig:nsdynamics}. Lines connect filled circles where calculations are performed. Final droplet volume from {\bf A} and {\bf B} (circles) agree with the NUPACK calculation when globally rescaled.
Each trace is an average of two distinct locations in one reaction chamber.
}
\end{figure*}

\section{Near-equilibrium condensation with purified RNA linkers}

As a first step in investigating the RNA-driven condensation of nanostars, we study nanostar phase separation mediated by purified RNA linkers that are equilibrated after a thermal quench.
RNA linkers are produced by {\it in vitro} transcription reactions in the absence of nanostars.
The resulting reaction mixture is treated with DNase I (New England Biolabs) to remove the DNA template, and the RNA is purified with a spin column (NEB Monarch RNA Cleanup Kit) to remove polymerase, excess ribonucleotides and incomplete transcripts ($\lesssim \! 15$~nt); see Supplemental Fig.~S1 for transcribed linker purity assay.
The resulting purified RNA linkers are resuspended in transcription reaction buffer (without RNAP), then added to different samples composed of nanostars (nanostar concentration $c_{NS} = 28~\mu$M) at varied nanostar/RNA stoichiometry. 
Samples are incubated for 24 hours at $37~^\circ$C to ensure proper equilibration, including allowing any condensed phase present to grow to a large enough size to be visualized with a light microscope (see Fig.~\ref{fig:nsdynamics}A); images are acquired with an epifluorescent microscope with a temperature-controlled stage set to $37^\circ$C.

Microscopy images show that the volume fraction of equilibrated RNA-linked nanostar droplets, at constant nanostar concentration $c_{NS} = 28~\mu$M, is non-monotonic with RNA concentration $c_{RNA}$ (Fig.~\ref{fig:nsdynamics}).
Condensed droplets do not appear at $c_{RNA}=2~\mu$M, but do appear at $6~\mu$M, implying that for this $c_{NS}$ the dilute phase boundary is between those values $2~\mu$M $< c_{dilute}< 6~\mu$M.
For $c_{RNA}>c_{dilute}$, RNA-nanostar droplets form, and the total condensed volume is linearly proportional with RNA concentration until $c_{RNA} \approx 2 c_{NS}$.
We find the peak of total droplet volume occurs for transcript concentration $c_{RNA_{max}}\approx 2c_{NS} = 56~\mu$M, and not $c_{RNA_{max}} = c_{NS} = 28~\mu$M, which implies that each linker can only bind two nanostars at one time, and not four (Fig.~\ref{fig:nsdynamics}B Inset).

As $c_{RNA}$ increases beyond $2c_{NS}$, the total droplet volume decreases.
We attribute this to the passivation of nanostar bonds, which are oversaturated by RNA linkers.
Indeed, in the limit $c_{RNA}\gtrsim 4c_{NS}$, we would expect every nanostar bond to be occupied with an RNA linker, thus disallowing any singly-bound RNA from finding an open bond on a second linker, preventing condensation.
That the maximum droplet volume occurs at $c_{RNA} \approx 2 c_{NS}$ is consistent with a molecular geometry in which each RNA can only participate in a single nanostar-nanostar cross-link.

\section{Nanostar condensation driven by \textit{in situ} transcription}
We then investigate the production of phase-separated droplets from transcription reactions occurring {\it in situ}.
We mix nanostars, RNAP, rNTPs, and DNA templates and observe the dynamics of droplet formation with fluorescent microscopy.
The images show that droplets are liquid-like, as evidenced by the presence of coalescence events during condensation (see Supplemental Movie 1). 
However, the droplets do not appear until after a significant delay ($t_d\approx$45 min; Fig.~2).
The delay time is also an order of magnitude larger than the droplet appearance time after temperature quenches with purified RNA linkers for $c_{RNA} = 6-30~\mu$M (see Supplemental Material Fig~S4).

To investigate the mechanism behind this delay, we conduct experiments in identical conditions to quantify the dynamics of complete RNA transcription and its impact on droplet formation (Fig.~\ref{fig:nsdynamics}B).
We first verify that the transcript purity and synthesis rate are not affected by the presence of DNA nanostars (see Supplemental Material Fig.~S1 and S2); this result reinforces the specificity of the transcription reaction, which is likely assisted by the branched structure of the nanostar preventing efficient RNAP binding to the DNA arms.
The measurements show that RNA linker transcription is linear at early times ($\lesssim$100 min; Fig.~\ref{fig:nsdynamics}B), then, at late times, the reaction slows, and the total amount of RNA linker produced reaches a plateau value.
From this data and the imaging data (Fig.~\ref{fig:nsdynamics}A), we create a parametric plot of the total droplet volume as a function of the transcribed RNA concentration and compare it to the equilibrated experiment using purified RNA (Fig.~\ref{fig:nsdynamics}B inset). 
The result emphasizes the non-equilibrium nature of the condensation dynamics when RNA linkers are produced, as the curve shapes are dramatically different, with {\it in situ} transcription leading to a pronounced suppression of droplet volume for small RNA concentrations.
From this plot, we find that the RNA concentration when droplets appear from {\it in situ} transcription is $c_{RNA}(45~\text{min}) = 40 \pm 5~\mu$M, substantially larger than the concentration required for equilibrium condensation, $c_{dilute} \approx 5~\mu$M (Fig.~\ref{fig:nsdynamics}B).
This shows the delay is not caused by a lack of RNA linker.
Finally, at high RNA concentrations, the two plots converge, indicating that {\it in situ} transcription eventually leads to near-equilibrium RNA-linked nanostar condensates after a pronounced transient.

Further analysis of the dynamics of transcription implies that a significant amount of abortive (incomplete) transcripts are formed.
At the plateau, the total amount of rNTPs incorporated into complete RNA transcripts is substantially less than the initial rNTP concentration (yield $\approx 20\%$).
This could be due to a finite RNAP lifetime; however, reactions with different template concentrations proceed initially at different rates and plateau at different times, with faster initial rates correlated with earlier plateau times (see Supplemental Figure~S3).
If the RNAP lifetime were limiting, we would expect the same plateau time regardless of the initial transcription rate.
Instead, these observations are consistent with the transcription reaction being limited by the total rNTP amount but with a significant number of rNTPs being incorporated into abortive transcripts.
To probe the source of the delay time in the {\it in situ} transcription system, we carry out experiments quantifying total droplet volume for different nanostar and RNAP concentrations.
Increasing the nanostar concentration leads to both earlier droplet appearance and more condensed volume (see Fig.~\ref{fig:ns_rnap_var}A). 
That droplets appear earlier with increased $c_{NS}$ is consistent with prior work on nanostar condensation after thermal quenches, where the timescales of droplet growth to a size visible in a fluorescent microscope were found to be limited by Brownian motion and coalescence ~\cite{wilken2024nucleation}). 
In such a mechanism, we expect the droplet appearance time to scale as $t_d \propto (1/c_{NS})^{1/3}$. 
Thus, the variation of initial condensate dynamics in Fig.~\ref{fig:ns_rnap_var}A (time to reach half-plateau volume is $t_d \approx [90$, $75$, $60$]~min, for $c_{NS} = [7.5$, $15$, $28$]~$\mu$M) suggests that Brownian motion and coalescence dictate the coarsening dynamics of {\it in situ} transcription-driven condensates.
That the final droplet volume is proportional to the bulk nanostar concentration is consistent with the lever rule of equilibrium phase diagrams, indicating again that the droplet condensation has equilibrated at the late stages of the reaction.

Experiments varying RNA polymerase concentration $c_{RNAP}$ show that {\it in situ} transcription experiments can be affected by nanostar passivation, just as in the equilibrated experiments with purified RNA.
As shown in Fig.~\ref{fig:ns_rnap_var}B, the timing of the appearance of visible droplets is not strongly dependent on $c_{RNAP}$, but the total droplet volume, at late times, varies inversely with $c_{RNAP}$.
To investigate this inverse dependence, we carry out NUPACK calculations that predict the number of RNA-linked nanostar complexes as a function of $c_{RNA}$ and $c_{NS}$.
These calculations predict a non-monotonic behavior as $c_{RNA}$ increases at constant $c_{NS}$, with a maximum at $c_{RNA}\approx 2 c_{NS}$, as expected due to the RNA oversaturation on nanostar bonds discussed above.
Notably, our late-time measurements of {\it in situ} transcribed droplet volume at various $c_{RNAP}$ and $c_{NS}$ can be globally rescaled to agree with NUPACK predictions (Fig.~\ref{fig:ns_rnap_var}C).
This clarifies that the inverse dependence of droplet volume on $c_{RNAP}$ occurs because higher $c_{RNAP}$ drives RNA production past the optimal value, causing a decrease in droplet volume as a result of nanostar bond passivation. 
This analysis further emphasizes that the delayed condensation response for {\it in situ} transcription, seen in Fig.~2, is \emph{not} due to passivation by full-length linkers, as the plateau RNA linker concentration is near optimum linker-nanostar stoichiometry.

Overall, our analysis shows that, while nanostar condensate formation can indeed be driven by {\it in situ} transcription, the resulting dynamics differ strongly from those for nanostar condensation in equilibrated conditions with purified RNA linkers.
The most likely explanation for the delay in the appearance of transcription-driven, relative to equilibrated, condensates is transient interference by abortive RNA transcripts. 
Abortive transcripts outnumber and diffuse faster than complete transcripts, which allows them to temporarily block the hybridization of complete transcripts to nanostars, delaying condensate formation. 
A similar role of interference by abortive transcripts was seen in prior work on synthetic genetic networks~\cite{kim2011synthetic}.
However, those abortive transcripts contain little or none of the nanostar-specific binding domain; thus, we expect the complete transcripts to dominate and drive condensation at long times. 
This is consistent with the long-time merging of the trajectories of condensate volume for {\it in situ} transcription and equilibrated measurements, as seen in Fig.~\ref{fig:nsdynamics}B inset.

\section{Reaction regulation via compartmentalization}

\begin{figure*}
\includegraphics[width=17cm]{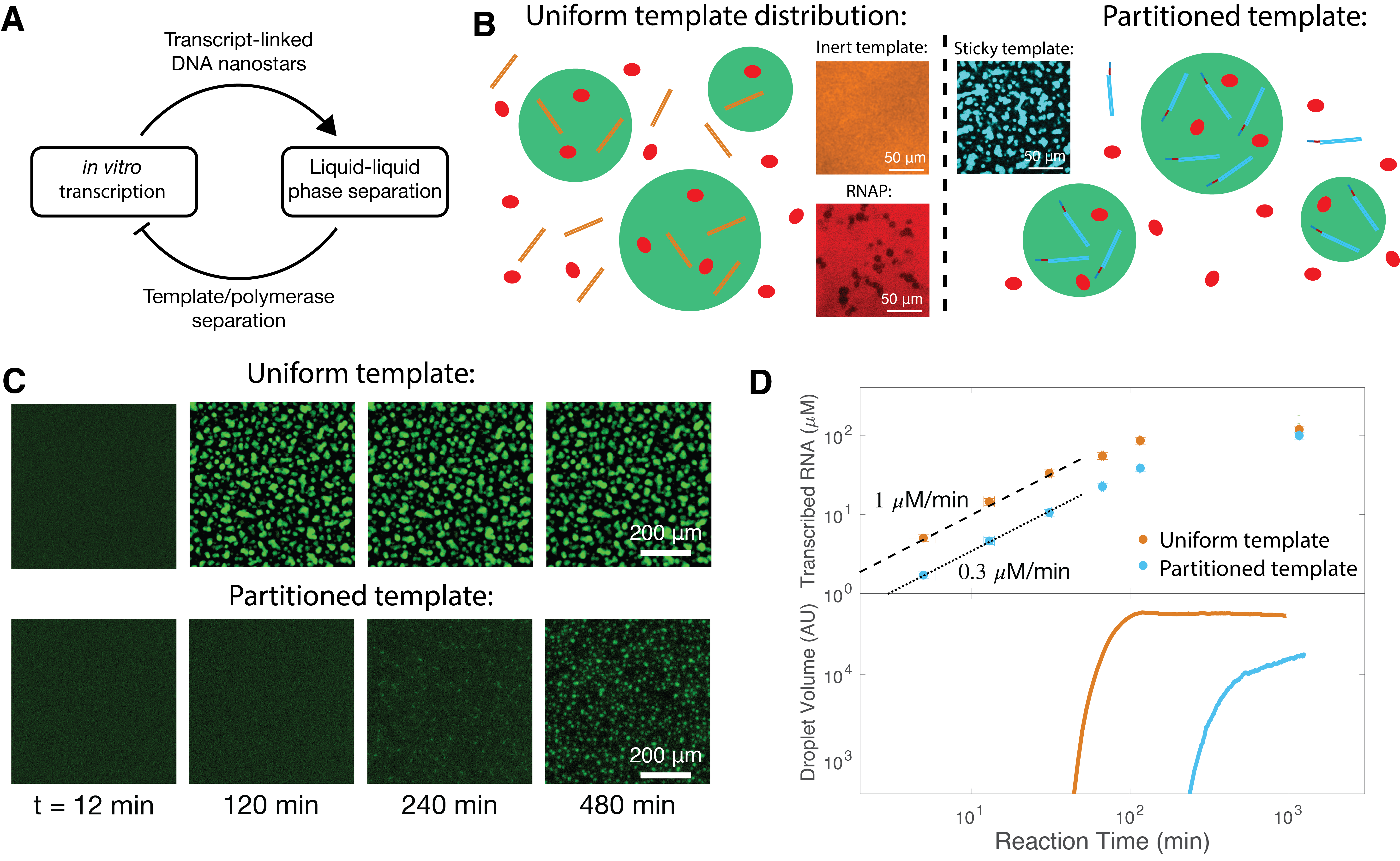}
\caption{\label{fig:txsilence} 
Activator/repressor control of RNA-linked nanostar droplets.
{\bf A} Control diagram of transcription-coupled liquid-liquid phase separation network. {\it In vitro} transcription triggers phase separation by linking DNA nanostars with RNA transcripts. Transcription reactions are then inhibited by separating DNA template from RNA polymerase.
{\bf B} Transcription repression mechanism: separating template from polymerase using nanostar droplets. Double-stranded ``inert" template concentration is equal inside and outside droplets, measured from fluorescent confocal microscopy ($P_{inert} = I_{in}/I_{out} = 1.0 \pm 0.1$).
A nanostar-binding sticky template is highly sequestered inside droplets ($P_{sticky} = I_{in}/I_{out} = 6 \pm 1$).
RNA polymerase (RNAP) is slightly excluded from droplets $P_{RNAP} = 0.7 \pm 0.1$. 
{\bf C} Fluorescent microscopy images of {\it in situ} transcription reactions comparing the condensation dynamics of the uniform template to the droplet-partitioned template.
Droplet formation is significantly delayed in the case where the template is separated from RNAP. 
{\bf D} Top: Gel electrophoresis quantification of the RNA concentration as a function of time reveals that both reactions proceed linearly, but the partitioned template reaction rate is suppressed by a factor of three. 
Bottom: Droplets produced by an inert template appear after $\sim 45$ min, whereas droplets produced by the partitioned template are substantially delayed ($\sim 300$ min). 
The factor of six increase in droplet appearance time is larger than the factor of three reduction in the transcription reaction rate.
}
\end{figure*}

We have demonstrated that distinct compartments of differing biochemical composition, i.e. dense regions of RNA/DNA inside droplets and dilute regions outside, can be generated by transcription reactions.
This implies that transcription rates may be tuned by localizing RNAP to a compartment with altered biochemical conditions.
One way to realize this is to induce RNA polymerase to partition largely into one compartment and the DNA template to the other, which would suppress the rate of transcription. 
Such a design would create a reaction feedback loop where a simple biochemical reaction ({\it in vitro} transcription) produces RNA, which in turn produces compartments, which then spatially sequester biomolecules in a manner that suppresses the reaction (see Fig.~\ref{fig:txsilence}A).


We first evaluate the local concentration of RNA polymerase and double-stranded DNA template inside droplets by fluorescent confocal microscopy. 
Separate reactions are performed such that droplets form in samples where a small fraction (10\%) of double-stranded DNA template (Cy5) or RNA polymerase (Alexa 647) are dyed.
T7 RNA polymerase is labeled using Succinimidyl (NHS) Ester (Thermo-Fisher Alexa Fluor 647 Microscale Protein Labeling Kit).
The fluorescent intensity of the double-stranded template indicates the template concentration is the same inside and outside droplets, with a partition coefficient $P_{inert} = I_{in}/I_{out} = 1 \pm 0.1$.
In addition, RNA polymerase is slightly excluded from nanostar droplets $P_{RNAP} = 0.7 \pm 0.1$ (see Fig.~4B).
Because the droplet volume fraction is small (10\%), the presence of droplets that only modestly exclude RNAP does not substantially impact the transcription reaction (see Supplemental Figure S2).

In order to sequester DNA templates inside droplets, we follow recent work~\cite{nguyen2019length}, which found that long DNA can be driven into the nanostar liquid phase by including single-stranded sequences that bind to nanostars.
The template-nanostar hybridization energy gain can overcome the entropic cost of confining long, semiflexible DNA within the nanostar liquid meshwork.
We design a sticky template that includes a nanostar-binding sequence and a toehold region (3 bases) so that template-nanostar hybridization is more favorable and, therefore, out-competes nanostar-nanostar binding (see Supplemental Material for complete template sequences).
In this case, fluorescent confocal images reveal that the template is highly partitioned into nanostar droplets, $P_{sticky} = 6 \pm 1$, while RNA polymerase remains excluded (Fig.~4B).

The stickiness of the template design substantially impacts transcription reaction kinetics. 
While the droplet phase is a minority of the total sample volume, we find that the sticky template substantially suppresses the transcription reaction rate.
The initial transcript production rate ($t<100$ min) is $0.30 \pm 0.05 \mu$M/min in the case where the template is partitioned, relative to the uniform template $1 \pm 0.1 \mu$M/min (see Fig.~4D).
At late times ($t>100$ min), the total amount of complete RNA transcripts in the partitioned case converges toward the inert case, suggesting that template partitioning does not substantially impact the abortive/productive transcription ratio.

The presence of two distinct biomolecular regions also produces substantially different condensation dynamics.
Reactions performed with the sticky template produce droplets, but only after a much longer time ($t_d \approx 200$ min) compared to transcription reactions with an inert template $t_d \approx 45$ min (Fig 4C).
The inert template produces enough RNA to link nanostars before droplets form so that droplets form all at once, and then the condensate volume is saturated for all times after $t = 100$ min.
In contrast, the sticky template produces micron-scale droplets much later. In addition, the total droplet volume continues to increase even after RNA production has ceased (at around \( t \gtrsim 1000 \) min). This suggests that the presence of the template within the droplets may inhibit coalescence and growth dynamics.

While we may have expected more complex reaction kinetics (e.g. one reaction rate before droplets form and a different/smaller rate after), template sequestration likely occurs on the timescale to produce droplets that are the linear size of the template (20 nm), which is a timescale too fast to resolve with our RNA quantification method.
Near-equilibrium condensation dynamics indicate that droplet formation is driven by spontaneous (spinodal) dynamics~\cite{wilken2024nucleation}, so we might observe less trivial reaction kinetics by positioning the nanostar hybridization energies and the RNA linker concentration such that condensation is driven by a nucleation and growth mechanism.
In addition, microscopy cannot resolve condensation dynamics on the scale of an individual template, so more interesting droplet condensation dynamics may be observed using techniques that can resolve the growth of small nanostar droplets, such as dynamic light scattering or fluorescence correlation spectroscopy.

\section{Conclusion}

We have developed a new model system to investigate reaction kinetics coupled to liquid-liquid phase separation, where DNA nanostars condense in the presence of transcribed single-stranded RNA linkers.
The condensation dynamics of this model system are distinctly non-equilibrium, even though RNA is continuously transcribed: droplets are only observed at much later times with a highly non-linear response relative to near-equilibrium condensation.
As expected, droplet formation is faster and produces more condensed volume for larger nanostar concentrations, but increasing RNA polymerase concentration, and therefore transcript concentration, surprisingly results in less condensed volume due to RNA-decorated nanostar passivation.
Finally, we design a model biomolecular activator/repressor network, where transcription activates droplet formation, producing spatial compartments of distinct biomolecular composition that inhibit transcription by separating the template from RNA polymerase.

The prominence of phase-separated regions in living cells has led to recent interest in creating condensates composed of artificial, engineered biomolecules, termed `synthetic cells'~\cite{polka2016building,adamala2024present}.
One way to control reaction pathways in these synthetic systems is by compartmentalizing biomolecules.
In particular, using nucleic acids, relative to, e.g. peptides, as a platform for artificial condensates provides extraordinary design flexibility due to the ease of varying sequence and the relatively straightforward relationship between sequence and structure~\cite{samanta2024dna}.
Recent works have demonstrated the creation of micron-scale compartments that have the ability to interface with biochemical reactions~\cite{deng2020programmable,malouf2023sculpting,saleh2020enzymatic} and perform simple computations~\cite{takinoue2023dna}.
Three recent works~\cite{fabrini2023co,stewart2023modular,udono2023programmable} demonstrated the formation of nucleic acid condensates based solely on single-stranded RNA produced by {\it in vitro} transcription reactions.
These studies use a different hybridization mechanism, exploiting RNA folding and tertiary interactions to create a branched structure of double-stranded hairpins that condense through tertiary (‘kissing loop’) interactions.
Our work on {\it in vitro} transcription-coupled condensation is driven by base-pairing and is highly complementary to recent work on {\it in vitro} transcription-coupled condensation driven by electrostatic coacervation~\cite{henninger2021rna}; the main advantage of our approach for synthetic cell reaction engineering is the high specificity afforded by sequence-specific hybridization.

Extensions of the work presented here may enable exploration of interesting, biomimetic, out-of-equilibrium behaviors, such as cycles of droplet growth and division, self-propulsion~\cite{demarchi2023enzyme,saleh2023vacuole}, core-shell structures~\cite{bergmann2023liquid}, size regulation~\cite{kirschbaum2021controlling,sastre2024size,gao2024controlling}, or droplet patterning~\cite{wilken2023spatial,banani2024active}. 
For example, it has been predicted that if the reaction rate of the production of phase-separating components is sufficiently different inside and outside of a droplet, then the droplet’s spherical shape becomes unstable, and a single droplet will divide into two daughter droplets~\cite{zwicker2017growth}.
In effect, the droplet morphology is modulated only by bulk properties, i.e. the reaction rates, as opposed to directly manipulating the interface~\cite{de2024dynamin}.
Phase-separated nucleic acid compartments coupled to enzymatic reactions might be an optimal system in which to explore these out-of-equilibrium behaviors, many of which have yet to be experimentally demonstrated.

\section{Methods}

\subsection{Sample Geometry/Imaging}
The sample chamber consists of a borosilicate glass microcapillary tube (Vitrocom) with interior dimensions: $300\mu $m height, $3$mm width, and $25$mm length.
Capillaries are internally coated with polyacrylamide~\cite{sanchez2013engineering} to keep droplets from sticking.
The droplets are imaged with a Nikon Ti-2 fluorescent microscope with a 10x objective. 
The full side-length of the acquired images is 1200 $\mu$m, and fluorescent images of droplets shown in the figures are 600 $\mu$m x 600 $\mu$m.
For imaging, 10\% of nanostars are modified by labeling one strand with a Cy3 fluorescent dye on the 3' end. 

\subsection{Reaction Conditions}
All reactions are performed with a custom transcription-nanostar buffer.
All transcription reactions contain 20mM total rNTPs (New England Biolabs) at a stoichiometry that matches the transcribed RNA sequence (8 mM GTP, 6.5 mM ATP, 4 mM UTP, 1.5 mM CTP).
Magnesium Acetate is added at a concentration of 26 mM, 6 mM in excess of the initial rNTP concentration.
20,000 kDa molecular weight Poly(ethylene glycol) from Polysciences is added at 2\% 
Dithiothreitol is added to a final concentration of 5 mM to stabilize RNA polymerase.
All samples are buffered with 10~mM Potassium Phosphate (pH = 6.9).
Potassium Acetate is added at 200~mM to increase the hybridization energy of RNA linkers to DNA nanostars.
Samples are prepared on ice, loaded into capillaries, and placed on a temperature-controlled microscope stage preheated to 37 $^\circ$C to image condensate volume throughout the reaction. 

\section{Image Analysis}

The total condensed volume is extracted from epifluorescent images of nanostar droplets by summing image intensities.
For all images, the images are corrected by flattening to correct for spatial inhomogeneity of the LED illumination by dividing by the intensity profile of the nanostar gas at the beginning of the reaction ($I_{profile}$).
Then, the dilute phase fluorescent intensity, estimated by the low-intensity peak in the histogram of pixel intensities at late times ($\approx 300$ min), is subtracted from all images ($I_{dilute}$).
Finally, the pixel intensities of the resulting images $I = (I_{raw}-I_{dilute})/I_{profile}$ are summed, which corresponds to the total fluorophore content and, therefore nanostar quantity in the dense liquid droplets.

\section{RNA quantification}

Transcribed RNA linker concentration is quantified by Urea denaturing polyacrylamide gel electrophoresis.
20 $\mu$L samples are incubated at 37 $^\circ$C inside a 100 $\mu$L PCR tube for all reactions.
At various times throughout the reaction, 2 $\mu$L of the sample volume is extracted, mixed with 100 $\mu$L DI water, and incubated at 70 $^\circ$C to denature the polymerase.
2 $\mu$L of the resulting dilution is mixed with 5 $\mu$L of 8M Urea and loaded into a 10\% polyacrylamide gel and run at 4mA (Biorad mini).
To quantify RNA linker length, the reaction mixtures are run alongside Low Range ssRNA Ladder and microRNA Marker (both from New England Biolabs).
To calibrate the RNA quantity, each gel is run with 25ng of purified DNA with the same sequence as the RNA linker. 
Gels are imaged with the Biorad GelDoc Go System and analyzed by custom code that individually integrates the peak in the intensity for each lane and subtracts the background fluorescence.
Conversion of gel intensity to concentration is done by comparing the same analysis on the lane containing 25ng DNA linker and correcting for the reaction dilutions (see Supplemental Figure 1). 

\begin{acknowledgments}
We thank William R. Jacobs for useful discussions. This work was supported by the W.M. Keck Foundation. 
\end{acknowledgments}

\bibliography{Tx_bib}

\end{document}




\title{Supplemental Material: Condensation dynamics and transcriptional control of a model biomolecular activator/repressor network}%
\author{Sam Wilken}
\affiliation{%
Materials Department, University of California, Santa Barbara, California 93106, USA
}%
\affiliation{%
Physics Department, University of California, Santa Barbara, California 93106, USA
}%
\author{Gabrielle R. Abraham}
\affiliation{%
Physics Department, University of California, Santa Barbara, California 93106, USA
}%
\author{Omar A. Saleh}
\affiliation{%
Materials Department, University of California, Santa Barbara, California 93106, USA
}%
\affiliation{%
Physics Department, University of California, Santa Barbara, California 93106, USA
}%

\date{\today}

\maketitle

\section{RNA-binding Nanostar Sequences}

\begin{tabular}{|l|l|} 
\hline
Strand 1 &
{\scriptsize
\texttt{\textbf{\textcolor{Cyan}{ACCTTCC}}\textbf{\textcolor{orange}{TGC}}GCGCGGCTCGCATTCCCGCGTTGGCAATGGGCCTGGCGGCGCT\textbf{\textcolor{Red}{TTCCAGA}}}} \\
\hline
Strand 2 &
{\scriptsize \texttt{GCGCCGCCAGGCCCATTGCCTTGCGGAGAACTGCCCGCGCGCT\textbf{\textcolor{Red}{TTCCAGA}}}}\\ 
\hline
Strand 3 &
{\scriptsize\texttt{\textbf{\textcolor{Cyan}{ACCTTCC}}\textbf{\textcolor{orange}{TGC}}GCGCGGCCGACGAGCGAGCGTTCGCGGGAATGCGAGCCGCGC}} \\ 
\hline
Strand 4 &
{\scriptsize \texttt{GCGCGCGGGCAGTTCTCCGCTTCGCTCGCTCGTCGGCCGCGC}} \\
\hline
Dyed Strand 4 &
{\scriptsize \texttt{/5Cy3/GCGCGCGGGCAGTTCTCCGCTTCGCTCGCTCGTCGGCCGCGC}} \\
\hline
\end{tabular}
\\ \\
Red and cyan regions correspond to nanostar sticky ends that bind transcribed RNA. Black sequences make up the core of the nanostar. Orange is a toehold region that only binds to DNA template to out-compete RNA-nanostar hybridization. Nanostar sequences were synthesized by IDT with standard desalted purification, except dyed strands, which are HPLC purified.

\section{DNA template sequences}

\begin{tabular}{ |l| } 
\hline
Nanostar-binding Template \\
\hline
{\scriptsize \texttt{/5Cy5/TTTTT\textbf{\textcolor{orange}{GCA}}\textbf{\textcolor{Cyan}{GGAAGGT}}\textbf{\textcolor{Red}{TCTGGAA}}ATAT\textbf{\textcolor{violet} {TAATACGACTCACTATA}}GGGATAAT\textbf{\textcolor{Cyan}{GGAAGGT}}\textbf{\textcolor{Red}{TCTGGAA}}GTCA\textbf{\textcolor{Cyan}{GGAAGGT}}\textbf{\textcolor{Red}{TCTGGAA}}}} \vspace{-.5cm} \\
{\scriptsize \hspace{4.79cm} \texttt{ATAATTATGCTGAGTGATATCCCTATTACCTTCCAAGACCTTCAGTCCTTCCAAGACCTT}}\\
\hline
Inert Template \\
\hline
{\scriptsize \hspace{2.925cm} \texttt{/5Cy5/TTTTTTAT\textbf{\textcolor{violet} {TAATACGACTCACTATA}}GGGATAAT\textbf{\textcolor{Cyan}{GGAAGGT}}\textbf{\textcolor{Red}{TCTGGAA}}GTCA\textbf{\textcolor{Cyan}{GGAAGGT}}\textbf{\textcolor{Red}{TCTGGAA}}}} \vspace{-.5cm} \\
{\scriptsize \hspace{4.79cm} \texttt{ATAATTATGCTGAGTGATATCCCTATTACCTTCCAAGACCTTCAGTCCTTCCAAGACCTT}}\\
\hline
\end{tabular}
\\ \\
Red and cyan regions correspond to nanostar-binding regions of the nanostar-binding template and the transcribed RNA. Orange region is a toehold sequence that increases template-nanostar hybridization energy. The purple region is the T7 RNA polymerase promoter sequence. Template sequences were synthesized by IDT with HPLC purification.

\bibliography{Nanostar_Hyperuniformity}

\newpage

\begin{figure*}
    \centering
    \includegraphics[width=\textwidth]{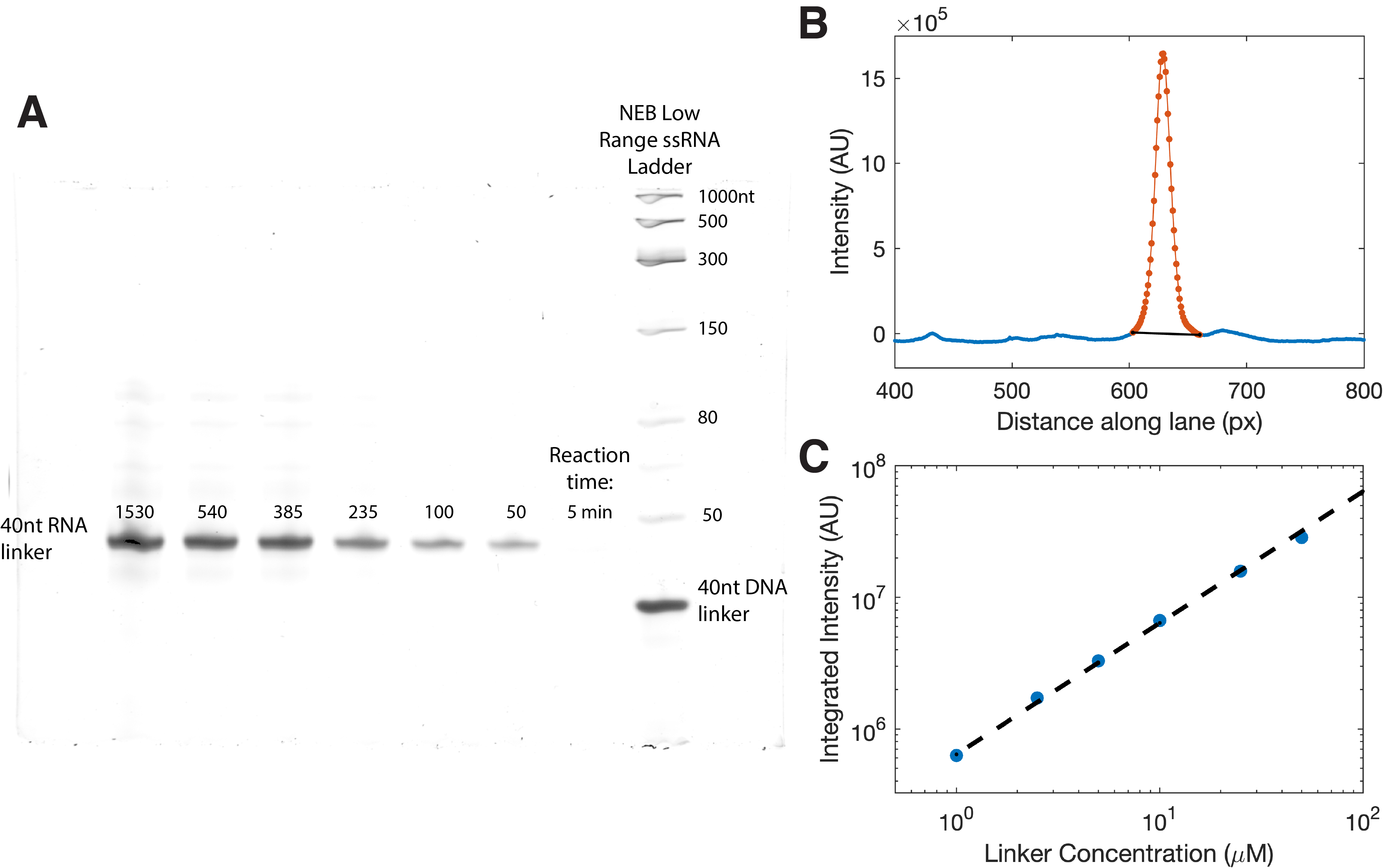}
    \caption{ \textbf{Gel electrophoresis RNA quantification}. {\bf A} Fluorescent image of a denaturing polyacrylamide gel containing a representative transcription reaction mixture, containing nanostars, at different times. At each time point, 2~$\mu$L of the reaction mixture is extracted, treated with DNase I (NEB) for 15 minutes, heated to $75^\circ C$ to denature the enzymes, and then held on ice. A single band of the target RNA linker (40 nucleotides) is found at all times, and more linker is produced as the reaction proceeds. For calibration, 25ng of a 40nt DNA linker is run along with NEB Low Range ssRNA Ladder to ensure proper length of the RNA linker. {\bf B} A representative intensity profile of one lane of the gel (here: 540 min). The area between the orange and the black indicates the total summed fluorescence. {\bf C} RNA linker concentration calibration. Different masses of RNA linkers are run on the gel between 0.5~ng and 25~ng (corresponding to 1 $\mu$M and 50 $\mu$M in transcription reactions) and compared to the mass as measured by UV-Vis spectrophotometer (Thermo Scientific NanoDrop One). Gel quantification of RNA linker is linear (dashed line) up to 25ng (or 50 $\mu$M in reaction mixtures before dilution).}
    \label{fig:enter-label}
\end{figure*}

\begin{figure*}
    \centering
    \includegraphics[width=.6\textwidth]{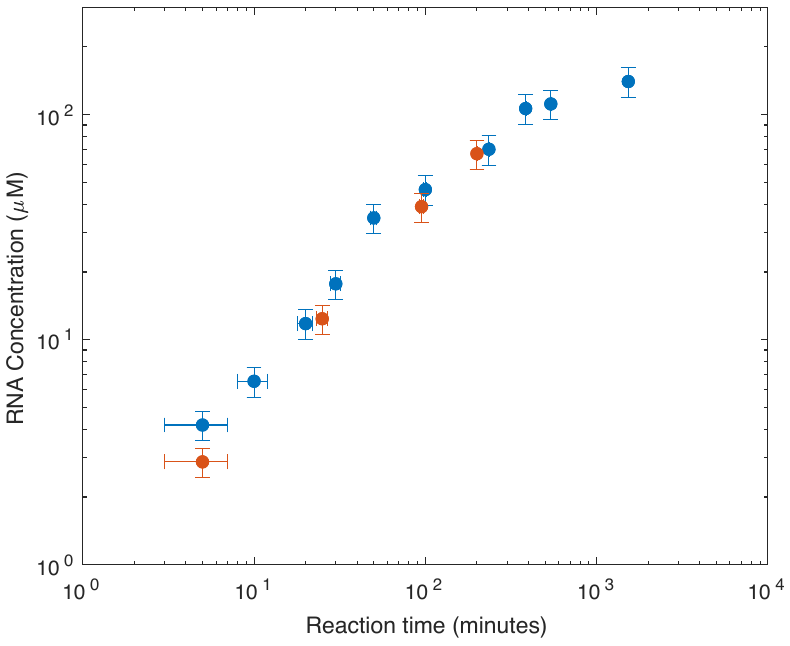}
    \caption{ \textbf{RNA production with and without nanostars}. The concentration of transcribed RNA, measured by gel electrophoresis, for reactions in the presence (orange) and absence (blue) of nanostars ($c_{NS} = 28~\mu$M). Both reactions were conducted with the inert template. Even though nanostars are designed to hybridize to transcribed RNA, the transcription reaction is not impacted by nanostars. The same data is presented in monochrome in the main text Fig.~2.
    }
\end{figure*}

\begin{figure*}
    \centering
    \includegraphics[width=.6\textwidth]{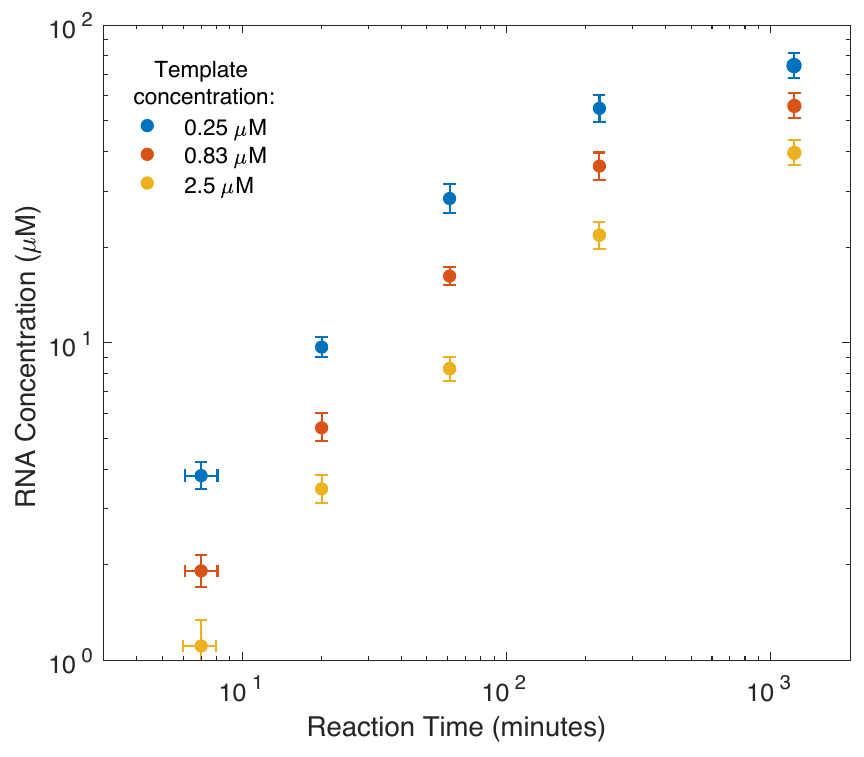}
    \caption{ \textbf{RNA production with varied template concentration}. The concentration of complete RNA transcripts, measured by gel electrophoresis, is plotted as a function of reaction time for samples with different DNA template concentration $c_{temp}=0.25~\mu$M (blue), $0.83~\mu$M (red), $2.5~\mu$M (yellow), in the absence of nanostars. All three reactions synthesize RNA linearly and then plateau as ribonucleotide concentration decreases. 
    By fitting to the form, $c_{RNA}(t) = c_\infty (1 - e^{-t/\tau_{RNA}})$ with fit parameters $c_{\infty} = [70 \pm 20~\mu$M, $55 \pm 15~\mu$M, $40 \pm 10~\mu$M] and $\tau_{RNA} = [150 \pm 40~$min, $200 \pm 50~$min, $280 \pm 60~$min] for $c_{temp} = [0.25~\mu$M, $0.83~\mu$M, and $2.5~\mu$M].
    }
\end{figure*}

\begin{figure*}
    \centering
    \includegraphics[width=\textwidth]{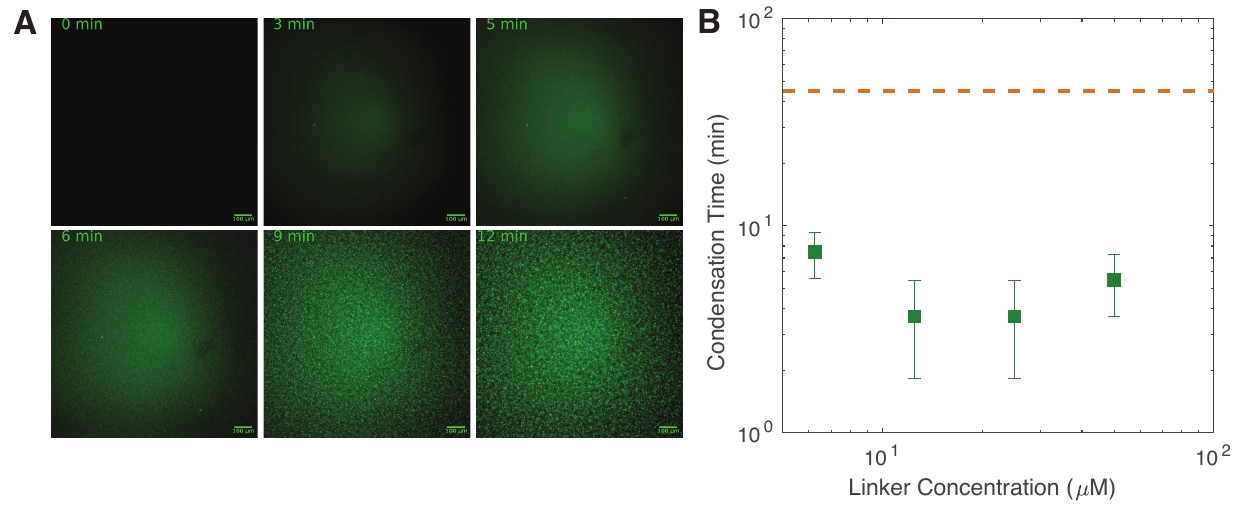}
    \caption{ \textbf{Near-equilibrium Condensation Time}. {\bf A} Microscopy images of the growth of an equimolar mixture of nanostars ($c_{NS} = 28~\mu$M) and purified RNA linkers ($c_{Link} = 28~\mu$M) after a temperature quench into the coexistence regime ($T = 37^\circ$C). The sample is prepared at $T = 50^\circ$C. Time $t=0$ is defined as the time when the sample temperature decreases below the melting temperature $T_m \approx 43^\circ$C. Droplets appear between time $t=5-6~$min (images are acquired at a rate of one per minute). Buffer conditions are identical to transcription reaction mixtures, without RNA polymerase and ribonucleotides. {\bf B} Quenches are performed for varied linker concentrations (at constant nanostar concentration $c_{NS} = 28~\mu$M). The condensation time is plotted as a function of linker concentration (green squares), we estimate the error as the uncertainty in $t=0$, approximately 2 minutes for all samples. Near-equilibrium condensation for all tested linker concentrations ($c_{link} = 6.25 - 50~\mu$M) is significantly faster than the droplet formation for linkers transcribed {\it in situ} (orange line) from main text Figure 2B.
    }
\end{figure*}

\begin{figure*}
    \centering
    \includegraphics[width=.7\textwidth]{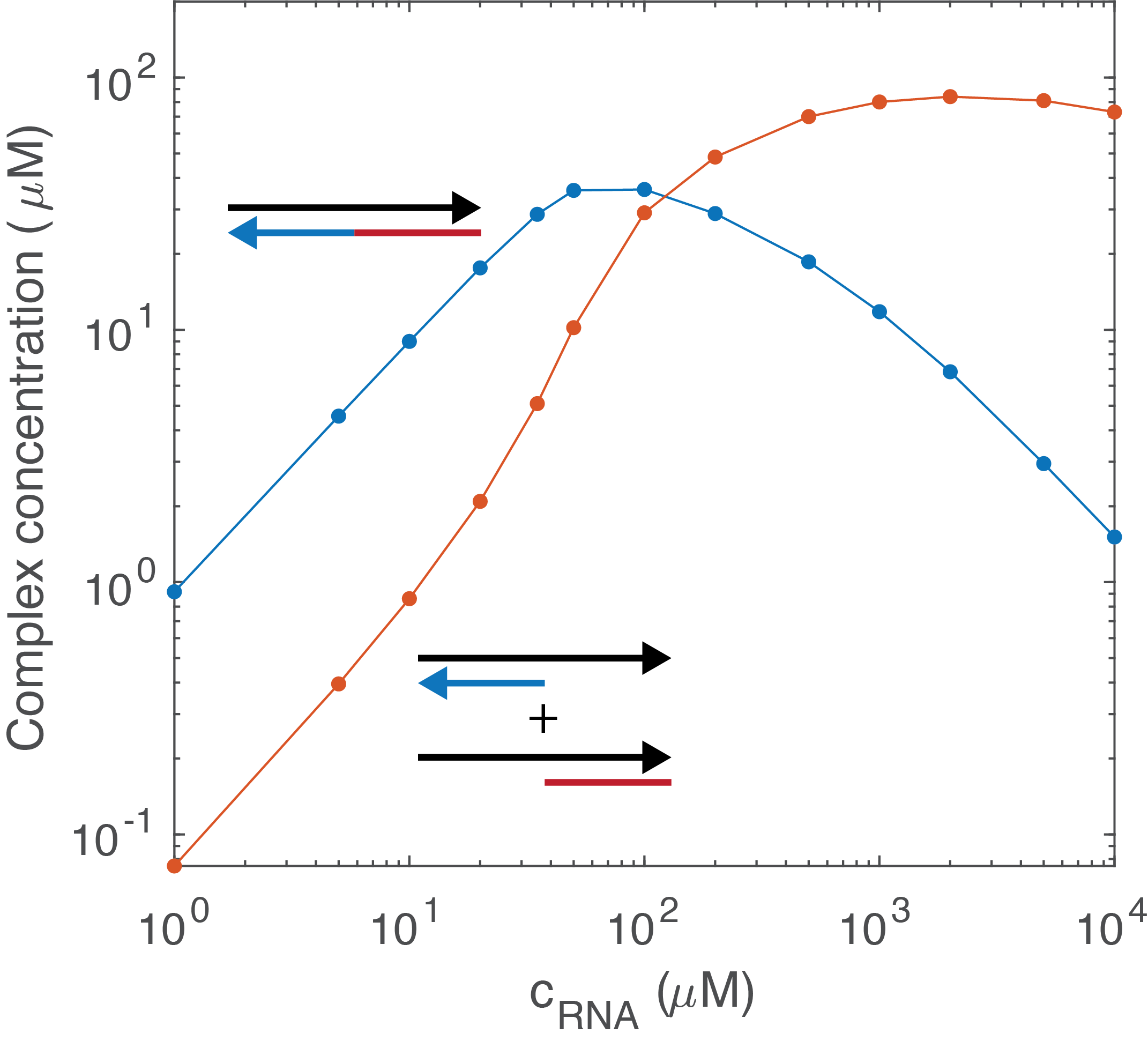}
    \caption{ \textbf{Equilibrium linker-nanostar complexation}. 
    NUPACK calculation of the equilibrium concentrations of linker-nanostar complexes for constant $c_{NS} = 28~\mu$M.
    The full nanostar-crosslinking complex is preferred for small linker concentration $c_{RNA}$. 
    For $c_{RNA} \gg 2c_{NS}$, RNA-linking nanostars is inhibited in favor of linkers only binding one nanostar sticky-end each.
    }
\end{figure*}